\documentclass[vecphys]{svmult}

\usepackage{makeidx}         
\usepackage{graphicx}        
\usepackage{multicol}        
\usepackage[bottom]{footmisc}

\makeindex             

\begin{document}

\title*{Extragalactic Stellar Astronomy}
\author{Miguel A. Urbaneja, Rolf-Peter Kudritzki \and
Fabio Bresolin\inst{1}}
\authorrunning{Urbaneja, Kudritzki \and Bresolin}
\institute{University of Hawaii Institute for Astronomy, 
2680 Woodlawn Drive, Honolulu HI\,96822, USA, 
\texttt{urbaneja@ifa.hawaii.edu, kud@ifa.hawaii.edu, bresolin@ifa.hawaii.edu}}
%
%
\maketitle

Despite their paucity, massive hot stars are real 
{\em cosmic engines} of fundamental importance in shaping our
Universe, from its very early stages up to its current appearance. 
Understanding the physics of massive stars is then a key issue for 
many relevant astrophysical phenomena. Their spectra provide constraints 
required for stellar and galactic evolution 
calculations, such as mass-loss rates, degree of chemical evolution at different
evolutionary stages, star formation history (through [$\alpha$/Fe] ratios) and
spatial distribution of different species. Probing the massive stellar population
of nearby galaxies by means of quantitative spectroscopy allows us to unveil
a wealth of information that will aid our current understanding of stellar and
galaxy evolution. In addition, blue luminous stars can be used as standard
candles for extragalactic distances up to 10 Mpc. 

Two main factors have contributed in recent years to mature the
field of extragalactic stellar astronomy. On the one hand, new observational
facilities, in particular ground-based large aperture telescopes equiped
with very efficient multi-object optical spectrographs, which make it possible to
collect data for a sensible number of stars in different galaxies. On the
other hand, enormous advances on model atmosphere techniques allow for the
first time the analysis of these large datasets by means of highly 
sophisticated and detailed models. In this contribution, we present a brief 
overview of recent steps we have undertaken in this exciting research field. 

\section{Quantitative spectroscopy of blue supergiant stars}
\label{urb_sec_two}

\begin{figure}
\centering
\includegraphics[height=0.9\textwidth,angle=90]{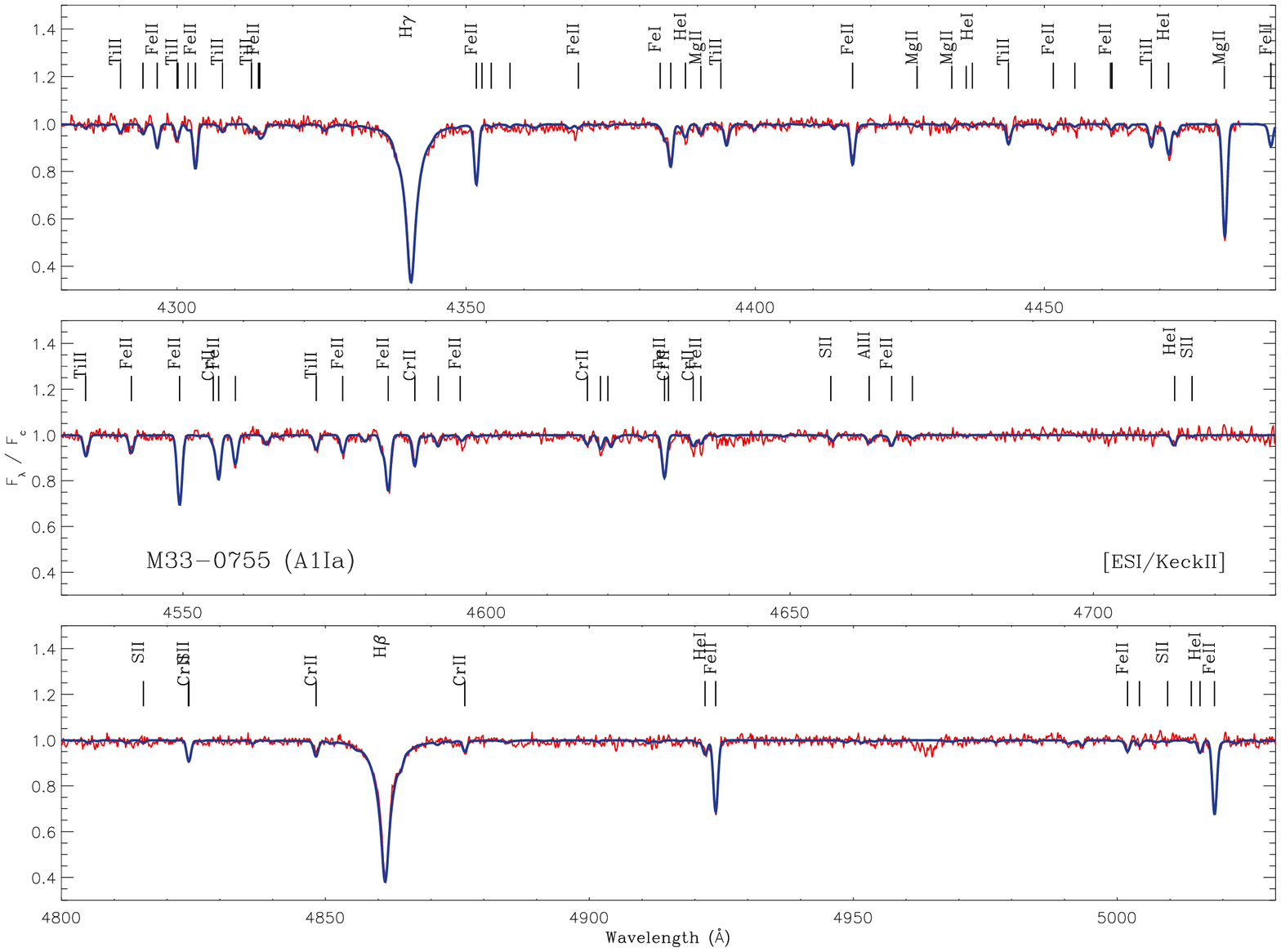}
\caption{Comparison of selected parts of the observed (ESI/KeckII,
R$\sim$10000) spectrum
of the early A supergiant M33-0755 and the best model. Most prominet spectral
features are identified.}
\label{urb_fig1}       
\end{figure}

The information about the physical properties of extragalactic massive stars 
is obtained through the comparison of the observed spectra with synthetic model 
atmospheres. In contrast to cooler spectral types, the physical processes in 
the atmospheres of blue supergiant stars are basically dominated by non-LTE conditions, 
due to the huge radiation field and the low density environment. As
previously quoted, the most recent generation of model atmospheres (\cite{hillier1998},
\cite{puls2005}, \cite{pauldrach2001}) are able to cope with the tremendous
challenge posed by these 
conditions in the presence of supersonic outflows (the stellar wind). The application
of such models to the analysis of samples of extragalactic stars, generally observed at lower
spectral resolutions, requires first a detailed study of stars in the Milky Way and nearby
galaxies to characterize the models and to assess our ability of reproducing the
observations (\cite{urbaneja2005b}, \cite{norbert2006}). To illustrate the quality than can be achieved, we
show in Fig. \ref{urb_fig1} selected parts of the observed spectrum
(ESI/Keck) of an early A Sg in M\,33 and the best fitting model.

Observation of stars in galaxies beyond the Local Group implies the use of low spectral
resolution (\cite{bresolin2001}, \cite{bresolin2002}). However, as we have shown in recent 
years (\cite{urbaneja2003}, \cite{urbaneja2005a}, \cite{bresolin2006}, \cite{evans2007}), it is 
possible to get information from these objects even at these low resolutions, provided that the
signal-to-noise is high enough (about $\sim$50). In the case of OB supergiants ($\sim$O9I--B3I) it 
is possible to follow the classic
techniques used with high-res data since the fundamental diagnostic lines are still available. In
the case of BA supergiants ($\sim$B4I--A3I), we have developed an alternative technique that
enables the analysis of these spectral types even when it is impossible to use the classic
methods (\cite{rolf2007}). 

\section{Blue supergiants' applications}

\subsection{Chemical abundances in nearby galaxies}

To date, studies of the spatial distribution of chemical species in galaxies have been mainly based
on oxygen abundances in H~{\sc ii} regions. As an alternative, and in some cases complementary, method, 
oxygen abundances can be determined from B- and A-type supergiant stars, by means of a solid and
self-consistent methodology, based on detailed analyses of the atmospheres of such
stars. The optical spectra of B- and A-type supergiants are rich in metal absorption 
lines from several elements (C, N, O, Mg, Al, S, Si, Ti, Fe, among others). As young 
objects they represent probes of the current composition of the inter-stellar 
medium (except for those species that are affected by the evolution 
of the star: C and N), hence these objects can be used to trace the present day abundance pattern 
in galaxies, with the ultimate goal of recovering its chemical and dynamical evolution history. 

We would like to stress that not only oxygen abundances, but a large number of chemical
species, in particular Fe-group elements (see Fig. \ref{urb_fig1}), are available through stellar 
spectroscopy. In our
recent work we have demonstrated how quantitative spectroscopy of blue supergiants can render 
accurate and relevant information about the spatial distribution of
different species in several  
nearby galaxies: WLM (\cite{bresolin2006}), IC\,1613 (\cite{bresolin2007}), NGC\,3109 (\cite{evans2007}), 
M\,33 (\cite{urbaneja2005b}) and NGC\,300 (\cite{urbaneja2005a}). 
For the Local Group galaxies studied so far, our results are consistent with nebular abundances obtained by the application of 
the direct method, while in the more distant NGC\,300 ($\sim$2 Mpc), our results support some empirical 
calibrations, while ruling out some others.

\subsection{The FGLR and distances in the local Universe}

The best established stellar distance indicators, such as Cepheids, RR Lyrae and RGB stars, suffer from
two major problems, extinction and metallicity dependence, both of which are difficult to
determine for these objects with sufficient precision. In order to improve distance determinations
in the local Universe and to assess the influence of systematic errors there is a need for alternative
distance indicators, which are at least as accurate but are not affected by uncertainties arising
from extinction or metallicity.

Blue supergiants are ideal objects for this purpose, because of their enormous
intrinsic brightness, which makes them available for accurate quantitative spectroscopic studies even
far beyond the Local Group. Quantitative spectroscopy allows us to determine stellar parameters and thus
the intrinsic spectral energy distribution, which can be used to measure reddening and the extinction law.
In addition, metallicity can be derived from the spectra.  

Theoretical calculations predict that massive stars evolve from the Main Sequence, 
in their way to the red, at almost constant luminosity and mass
(\cite{rolf2003}). During this brief transitionary period, the ratio of the
effective temperature and the effective gravity remains constant. The relationship 
between the mass and luminosity of
supergiants means that the luminosity of blue supergiants in this phase can
be inferred from measurements of the effective temperature and effective
gravity alone.  This means that spectroscopic observations of blue
supergiants in other galaxies may be used, through the Flux-weighted Gravity
-- Luminosity Relationship, FGLR, to determine distances to these galaxies.  This
technique is robust, we have shown how observations of at least 10--15 blue
supergiants may provide a distance modulus with an uncertainty of 0.1
mag (\cite{rolf2003}).  We have also shown, from observations of blue
supergiants in NGC\,300, that the photometric variability has negligible
effect on the distances determined through the FGLR  (\cite{bresolin2004}). The advantage 
of the FGLR-technique is the fact that individual metallicity, reddening and extinction 
can be determined for each star directly from
spectroscopy combined with photometry. 

%
%

%
%
%
%

%
%



\printindex
\end{document}